\documentclass[12pt]{iopart}
\usepackage{iopams}
\usepackage{graphicx}

\newcommand{\eqn}{\begin{eqnarray}}
\newcommand{\feqn}{\end{eqnarray}}
\newcommand{\beq}{\begin{equation}}
\newcommand{\eeq}{\end{equation}}
\newcommand{\bes}{\begin{equation*}}
\newcommand{\ees}{\end{equation*}}
\newcommand{\smat}{\left( \begin{smallmatrix}}
\newcommand{\smct}{\end{smallmatrix}\right)}

\newcommand{\beqnl}{\begin{eqnarray}}
\newcommand{\eeqnl}{\end{eqnarray}}

\begin{document}

\title{Stimulated emission and Hawking radiation in black hole analogues}
\author{F.~Belgiorno$^{1,2}$ and S.L.~Cacciatori$^{3,4}$}

\address{$^1$ Dipartimento di Matematica, Politecnico di Milano, Piazza Leonardo 32, 20133 Milano, Italy}
\address{$^2$ INdAM-GNFM, Italy}
\address{$^3$ Department of Science and High Technology, Universit\`a dell'Insubria, Via Valleggio 11, IT-22100 Como, Italy}
\address{$^4$ INFN sezione di Milano, via Celoria 16, IT-20133 Milano, Italy}

\ead{\mailto{francesco.belgiorno@polimi.it}, \mailto{sergio.cacciatori@uninsubria.it}}

\begin{abstract}
Stimulated emission by black holes is discussed in light of the analogue gravity program. 
We first consider initial quantum states containing a definite number of particles, and then we take into account the 
case where the initial state is a coherent state. The latter case is particularly significant in the case where Hawking 
radiation is studied in dielectric black holes, and the emission is stimulated by a laser probe. We are particularly 
interested in the case of the electromagnetic field, for which stimulated radiation is calculated too. 
\end{abstract}

\maketitle

\section{Introduction}

Stimulated emission by black holes has is a longstanding topic in quantum field theory on a black hole background, 
which was taken into account since the very early studies in black hole evaporation \cite{wald,meisels}. See also 
\cite{audretsch,frolov}. Since the 
former analysis, it appeared that stimulated radiation is far from being of practical interest in the case of astrophysical 
black holes. Still, a relevant role for stimulated radiation is deserved in some attempts to solve 
the unitarity problem, because of the fact that stimulated radiation carries out information \cite{adami14,adami15}. 
We don't delve into the latter aspect.  
A different consideration for the same topic has to be deserved in the case of analogue gravity, because 
conditions where stimulated emission can play a very relevant role can be actually realized in labs. In particular, 
we are interested in dielectric black holes which are obtained as moving dielectric perturbations associated 
with strong laser pulses in nonlinear dielectric media via Kerr effect. Indeed, a further laser probe, of weak 
intensity (so not participating the Kerr effect) can be shot onto the dielectric perturbation, in such a way that 
a intense stimulated emission of pairs can occur. The conditions under which such stimulated emission is possible 
have been studied, both numerically and experimentally, by D.Faccio group in \cite{rubino-njp,rubino-sr,petev-prl}. 
We limit ourselves to recall that the original idea 
of the Kerr effect as a tool for studying Hawking radiation in dielectric black holes can be traced back to Ref. \cite{philbin}, 
and a series of papers on the subject appeared in the literature \cite{faccio-prl,belgiorno-prd,unshu,faccio-prl-ans,
rubino-njp,unshu-prd,prain,finazzi-carusotto-pra13,rubino-sr,finazzi-carusotto-pra14,petev-prl,PRD2015,jacquet}.\\

Herein, we take into account some theoretical aspects of this stimulation phenomenon. 
We first follow strictly the presentation given in \cite{audretsch}, which adopts the strategy of Bogoliubov transformations 
between IN and OUT states in a collapse situation. This strategy is well-known in black hole emission process since the seminal 
calculation by S.W.Hawking. As usual, a Fock space state with a definite number of particles is taken into account. 
We also provide a very simple deduction of the stimulated contribution to pair creation by means of 
thermofield dynamics formalism, which is particularly useful once more.\\
As a further matter of analysis, we take into consideration the case of a coherent state as initial state. This choice 
is, to some extent, on the opposite side with respect to the standard choice of a quantum state belonging to the Fock space, 
as it represents the best approximation to a classical state, to be compared with the eminently quantum nature of a 
Fock state. The coherent state stimulation appears of limited interest in the astrophysical black hole state, 
but is actually a very interesting topic in the case of dielectric black holes. Indeed, stimulated Hawking radiation 
can be obtained as described above, by means of a weak laser probe, whose nature of coherent light is well-known.  
This same strategy can be used also in the case of dielectric dispersive black holes, 
and we shall use it also for the standard ternary process which is at the root of Hawking radiation in the 
dielectric case.\\
We don't consider herein a full calculation for the full electromagnetic case, which appears to be very 
involved, and limit our considerations to some general properties we expect to be implemented also in the 
full calculation, which is deferred to future works. 
The conclusions that we can infer  for the standard ternary process $IN\to P+N$, where $IN$ stays for the 
input particle state, and $P$ and $N$ stay for the particle and antiparticle final states, 
are the expected ones: the created pairs are such that one emitted photon  is found in the 
P mode peak and the companion photon (antiparticle) is found in the N mode peak. The spontaneous contribution 
is unpolarized, as it should be due to thermality of the spontaneous radiation, whereas the stimulated one is 
suitably polarized, in such a way that a created photon is polarized in the same way as the stimulating particle or antiparticle state. As a consequence, given an IN state 
populated by particle states with a given polarization, one obtains a P mode peak and a N mode peak, all with the same polarization. 

\section{Stimulated emission and black holes}

We start by summarizing some calculations appearing in \cite{audretsch}. 
Let $\{f_i\}$ be a set of positive norm solutions for the Klein-Gordon equation in the initial state labeled by $IN$ (no black hole), and 
let us define $\{p_i\}$ as positive norm solutions available at infinity when a static black hole is present (label $OUT$). 
As known, to the latter 
set one has to join a further set of states $\{q_i\}$ which are not available to the distant observer (horizon states, label $H$) in order to 
get a basis for the solutions in presence of a static black hole. As both $\{f_i\}$ and $\{p_i\}\cup \{q_i\}$ represent a basis of solutions, 
one can e.g. write
\beqnl
p_i &=& \sum_j (\alpha_{ij} f_j+\beta_{ij} f^\ast_j),\\
q_i &=& \sum_j (\gamma_{ij} f_j+\eta_{ij} f^\ast_j),
\eeqnl
together with 
\beq
f_j=\sum_i (\alpha_{ij}^\ast p_i-\beta_{ij} p^\ast_i+\gamma_{ij}^\ast q_i-\eta_{ij} q^\ast_i).
\eeq
As to the (charged) field, we get
\beqnl
\phi (x)&=&\sum_i (a_i^{IN} f_i+b_i^{IN\ \dagger} f^\ast_i)\\
&=& \sum_i (c_i^{OUT} p_i+d_i^{OUT \dagger} p_i^\ast +g_i^{H} q_i+h_i^{H\ \dagger} q^\ast_i),
\eeqnl
where $IN,OUT,H$ label the states in the initial condition, the outgoing modes in presence of black hole and the 
horizon states respectively. As to the relations between creation-annihilation operators in the different bases, we are 
interested in 
\beqnl
c_i^{OUT} &=& \sum_j (\alpha_{ij}^\ast  a_j^{IN}-\beta_{ij}^\ast b_j^{IN\ \dagger} ),\\
d_i^{OUT\ \dagger} &=& \sum_j (\alpha_{ij}  b_j^{IN\ \dagger}-\beta_{ij} a_j^{IN} ),
\eeqnl
and conjugate ones.
The number of particles operator for the mode $k$ is 
\beqnl
N_k^{OUT} &=& c_k^{OUT\ \dagger} c_k^{OUT} \cr
&=& \sum_{jl}  (\alpha_{kj} \alpha_{kl}^\ast  a_j^{IN\ \dagger} a_l^{IN} -
\alpha_{kj}  \beta_{kl}^\ast   a_j^{IN\ \dagger} b_l^{IN\ \dagger}\cr
&&\hphantom{\sum}-\beta_{kj} \alpha_{kl}^\ast  b_j^{IN} a_l^{IN} 
+\beta_{kj}  \beta_{kl}^\ast b_j^{IN} b_l^{IN\ \dagger} ).
\eeqnl
Spontaneous pair creation occurs when 
\beq
< 0_{IN} | N_k^{OUT} |0_{IN} > = \sum_j |\beta_{kj}|^2 
\eeq
is different from zero. Stimulated emission occurs if there are particles (and/or antiparticles) in the initial state: let 
\beq
|\psi_{IN}> \in {\mathcal{F}},
\eeq
i.e. let the initial state belong to the Fock space ${\mathcal{F}}$ and be an eigenstate of the number operator 
$N^{IN}$. Then, it is straightforward to show that 
\beqnl
< \psi_{IN} | N_k^{OUT} |\psi_{IN} >& =& \sum_j |\beta_{kj}|^2 + \sum_j |\alpha_{kj}|^2 <n_j^{IN}> \cr
&+& 
 \sum_j |\beta_{kj}|^2 <\bar{n}_j^{IN}>, 
\eeqnl
where $n_j^{IN}$ stays for the number of particles in the $j$-state IN and 
$\bar{n}_j^{IN}$ stays for the number of antiparticles in the $j$-state IN. As it is evident, 
the first contribution is associated again with the spontaneous pair creation, whereas 
the two further contributions are associated with the particle and antiparticle content of the 
initial state, and represent stimulated pair creation contributions.\\
For the antiparticle number operator, we have 
 \beqnl
\bar{N}_k^{OUT} &=& d_k^{OUT\ \dagger} d_k^{OUT} \cr
&=& \sum_{jl}  (\alpha_{kj} \alpha_{kl}^\ast  b_j^{IN\ \dagger} b_l^{IN} -
\alpha_{kj}  \beta_{kl}^\ast   b_j^{IN\ \dagger} a_l^{IN\ \dagger}\cr 
&&-\beta_{kj} \alpha_{kl}^\ast  a_j^{IN} b_l^{IN} 
+\beta_{kj}  \beta_{kl}^\ast a_j^{IN} a_l^{IN\ \dagger} ).
\eeqnl
so that 
\beq
< 0_{IN} | \bar{N}_k^{OUT} |0_{IN} > = \sum_j |\beta_{kj}|^2, 
\eeq
and 
\beqnl
< \psi_{IN} | \bar{N}_k^{OUT} |\psi_{IN} > &=& \sum_j |\beta_{kj}|^2 + \sum_j |\alpha_{kj}|^2 <\bar{n}_j^{IN}> \cr
&+& 
 \sum_j |\beta_{kj}|^2 <n_j^{IN}>.
\eeqnl
In the case of diagonal Bogoliubov transformations, we obtain
\beq
< \psi_{IN} | N_k^{OUT} |\psi_{IN} > = |\beta_{k}|^2 + |\alpha_{k}|^2 <n_k^{IN}> + 
 |\beta_{k}|^2 <\bar{n}_k^{IN}>, 
 \label{diag-number}
\eeq
and 
\beq
< \psi_{IN} | \bar{N}_k^{OUT} |\psi_{IN} > = |\beta_{k}|^2 + |\alpha_{k}|^2 <\bar{n}_k^{IN}> + 
 |\beta_{k}|^2 <n_k^{IN}>.
\eeq
It is worth noting that the above formulas hold true also in a more generic case where one passes 
from an IN state to a OUT one. Indeed, the horizon states do not affect the results for what concerns 
the number of particles measured at infinity.\\
According to the calculations in \cite{meisels}, which amount substantially to the ones obtained 
in \cite{audretsch} by means of a careful wave packet analysis in a collapse situation, one finds 
in the black hole case
\beq
< \psi_{IN} | N_\omega^{OUT} |\psi_{IN} > = \frac{1}{e^{\beta \omega}-1} +   \frac{e^{\beta \omega}}{e^{\beta \omega}-1}<n_\omega^{IN}> + 
\frac{1}{e^{\beta \omega}-1}  <\bar{n}_\omega^{IN}>, 
\eeq
and 
\beq
< \psi_{IN} | \bar{N}_\omega^{OUT} |\psi_{IN} > = \frac{1}{e^{\beta \omega}-1} +   \frac{e^{\beta \omega}}{e^{\beta \omega}-1}<\bar{n}_\omega^{IN}> + 
\frac{1}{e^{\beta \omega}-1}  <n_\omega^{IN}>, 
\eeq
where $\beta$ is the inverse black hole temperature. It is interesting to note that, when the stimulated effect dominates over the spontaneous one, and in absence of initial antiparticles, one obtains
\beq
< \psi_{IN} | N_\omega^{OUT} |\psi_{IN} >  \sim 
\frac{e^{\beta \omega}}{e^{\beta \omega}-1}<n_\omega^{IN}> .
\eeq
It is worth noting that stimulated emission can be calculated also by assuming as thermal state the Hartle-Hawking one, 
and by exploiting the relations between Hartle-Hawking state and Thermofield Dynamics as in \cite{israel}. The 
calculation is straightforward, and proceeds as follows. 
We recall that in the thermofield dynamics formalism one defines a
thermal state $|0(\beta)>$ characterized by an inverse temperature $\beta$. This state requires a 
doubling of degrees of freedom with respect to the standard quantum field theory at zero temperature, 
and the (unobservable) copy of the fictitious Hilbert space is characterized by operators with a 
tilde symbol.  The thermal state $|0(\beta)>$ 
is annihilated by suitable operators $a_l (\beta),\tilde{a}_l(\beta)$, 
$b_l(\beta),\tilde{b}_l(\beta)$ (and conjugated ones) which are labeled by a complete set of
quantum numbers $l$ and is related to ``standard''
annihilation-creation operators $a_l,\tilde{a}_l,b_l,\tilde{a}_l$ (and conjugated ones)
via a formally unitary transformation:
\beqnl
a_l &=& \cosh (\phi_{\omega_l})  a_l (\beta) + \sinh (\phi_{\omega_l}) \tilde{a}_l^{\dagger} (\beta),\\
\tilde{a}_l^\dagger &=& \sinh (\phi_{\omega_l}) a_l (\beta) +  \cosh (\phi_{\omega_l})   \tilde{a}_l^{\dagger} (\beta),
\eeqnl
and analogous for $b$-operators, with
\beq
\tanh (\phi_{\omega_l})= e^{-\beta \omega_l/2}.
\eeq
It holds
\beq
|0(\beta)> = e^{-iG} |0,\tilde{0}>,
\eeq
where 
\beq
G=\sum_l \log (\tanh (\beta \omega_l)) \left[(\tilde{a}_l a_l - \tilde{a}_l^\dagger a_l^\dagger)+
(\tilde{b}_l b_l - \tilde{b}_l^\dagger b_l^\dagger)\right].
\eeq
In the following, we indicate by HH the Hartle-Hawking state and by B the Boulware state in the case of Schwarzschild 
black hole, and $R$ and $L$ are the labels for the right and left region of the Kruskal diagram, as usual. $L$ is the label 
for unobservable states, of course. For simplicity, 
we consider a neutral scalar field (and then only $a$-modes appear). Moreover, we indicate with $|B>$ the state $|B,\tilde{B}>$. We get 
\begin{eqnarray}
|HH>=\exp(-i G) |B>&=&\exp \left[ \sum_\omega (\frac{1}{2} \log (1-\exp (-\frac{\pi}{2 \kappa} \omega))\right]\cr
&&\exp \left[   \exp (-\frac{\pi}{\kappa} \omega) 
\sum_{j}\left( a^{R \dagger}_{\omega j} a^{L \dagger}_{\omega j}    \right) \right]  |B>.
\end{eqnarray}
We have also
\begin{eqnarray}
|HH>=\frac{1}{Z(\kappa)} \prod_{\omega j} \left(\sum_{n_{\omega j}=0}^\infty \exp \left(-n_{\omega j} \frac{\pi \omega}{\kappa}
 \right) \right) |n_{\omega j} >^L  |n_{\omega j} >^R.
\end{eqnarray}
States $ |n_{\omega j} >^L, |n_{\omega j} >^R$ are Boulware states with  $n_{\omega j}$ particles. Killing observers in the R-region 
cannot measure $L$-states, so a trace over the latter ones is required. As a consequence, for the expectation value of an operator 
(observable) $A^R$, constructed by means of Boulware vacuum creation and annihilation operators, one obtains
\begin{equation}
<HH|A^R|HH>=\frac{1}{Z(\kappa)} \sum_n \exp (-\beta^\ast_H E_n) <n |A^R|n>,
\end{equation}
where $E_n = \frac{1}{K} \sum_{\omega j} n_{\omega j} \omega$, $\beta^\ast_H=\beta_H K$ ($K$ stays for the norm of the timelike Killing vector), 
$Z=\sum_n \exp (-\beta^\ast_H E_n)$. The relation with thermofield dynamics is then implemented, with the identification of the fictitious states with the 
states in the unaccessible $L$-region. In particular, we get
\beq
<HH|N_k^R |HH>=<HH|a^{R \dagger}_k a^R_k |HH>=\frac{1}{e^{\beta \omega_k}-1},
\eeq
as expected.\\

\noindent Thermofield dynamics formalism allows us to set out the following identifications: 
\beqnl
\alpha_k &=& \cosh (\phi_{\omega_k}),\\
\beta_k &=& \sinh (\phi_{\omega_k}),
\eeqnl
i.e. the Bogoliubov transformation is diagonal and, moreover, real valued. 
According to the definition in \cite{khanna}, we can introduce thermal number states 
as follows (for simplicity, we refer to a single mode of the field): 
\beq
|n(\beta)>=\frac{1}{\sqrt{n!}} (a^\dagger (\beta))^n |0(\beta)>, \quad \quad n=0,1,2,\ldots
\eeq
Stimulated emission can be obtained by calculating the mean value of $N_k,\bar{N}_k$ 
on thermal number states: we define 
\beq
<A_k>_{n,\bar{n},\beta}:=< n_k (\beta), \bar{n}_k (\beta)| A_k | n_k (\beta), \bar{n}_k (\beta)>,
\eeq
then we obtain 
\beqnl
<N_k>_{n,\bar{n},\beta} &=& 
\sinh^2 (\phi_{\omega_k}) + \cosh^2 (\phi_{\omega_k}) n_k (\beta) + \sinh^2 (\phi_{\omega_k}) \bar{n}_k (\beta),\\
<\bar{N}_k>_{n,\bar{n},\beta} &=&
\sinh^2 (\phi_{\omega_k}) + \sinh^2 (\phi_{\omega_k}) n_k (\beta) + \cosh^2 (\phi_{\omega_k}) \bar{n}_k (\beta),
\eeqnl
which correspond to the above mentioned results for stimulated emission by black holes. Note that
\beqnl
\sinh^2 (\phi_{\omega_k}) &=& \frac{1}{e^{\beta \omega_k}-1},\\
\cosh^2 (\phi_{\omega_k}) &=& \frac{e^{\beta \omega_k}}{e^{\beta \omega_k}-1}.
\eeqnl
Stimulated emission appears as the mean value of the number operator over an `initial' state 
which is a thermofield state with non-zero thermal particles/antiparticles. In a black hole context, 
we have non-zero Hartle-Hawking particles stimulating the emission process. A more direct physical intuition 
can be provided by the Unruh effect situation: non-zero particles as excitations of the Minkowski  vacuum 
stimulate the thermal emission for the Rindler observer. 

\section{Stimulated emission and initial coherent states}

In the previous section, we discussed what happens if the initial state is characterized by a definite (and finite) 
number of particles and/or antiparticles. These states are standard in quantum field theory calculations, 
but don't cover exhaustively the possibilities one could have in mind. For example, one could take into account 
coherent states, which correspond to field states as near as possible to classical states of the fields. 
This consideration of coherent states, which appears to be of purely speculative interest in the case of 
black holes, is instead of great interest in case of black hole analogues. Indeed, in the case of dielectric 
black holes which are generated by means of the Kerr effect, a dielectric perturbation traveling in the 
dielectric medium plays the role of the black hole, and is generated by a strong laser pulse. When a 
further weak laser probe is shot against the probe pulse, one can obtain stimulated emission. A formal 
way to describe the laser probe is by means of coherent states, which makes very strong and worthwhile of direct physical 
application the interest in this topic.\\
We recall that a coherent state is defined as an eigenstate of the annihilation operator. For example, if $\hat{c}$ is an annihilation 
operator, such that
\beq
\hat{c}|0> =0,
\eeq
and $[\hat{c},\hat{c}^\dagger]=1$, then a coherent state $|z>$ is such that 
\beq
\hat{c} |z>=z |z>,
\eeq
and also 
\beqnl
|z> &=& e^{-|z|^2/2} \sum_{n=0}^\infty  \frac{z^n}{n!} |n>\cr
&=& e^{z \hat{c}^\dagger-z^\ast \hat{c}} |0>,
\eeqnl
where $|n>=c^{\dagger n} |0>$ has norm $\sqrt {n!}$. 
We recall also that a Poissonian distribution of particles corresponds to the coherent state, 
and also that coherent states can be normalized (we assume that ours ones in the following are 
normalized) but not orthogonal each other, and form an overcomplete set of states. Moreover, in a field theory 
framework, we should write \cite{zuber}
\beq
|\eta> =e^{\int \frac{d^3 \pmb k}{(2\pi)^3 2\omega_k} \left( \eta (\pmb k) \hat{c}^\dagger (\pmb k) - 
 \eta^\ast (\pmb k) \hat{c} (\pmb k)\right)} |0>,
 \eeq 
with 
\beq
\int \frac{d^3 \pmb k}{(2\pi)^3 2\omega_k} | \eta (\pmb k) |^2<\infty.
\eeq
We recall that 
\beq
\hat{c} (\pmb k) |\eta> =  \eta (\pmb k) |\eta>.
\eeq
However, in the following, we shall keep the discrete notation, in order to simplify the presentation. 
Like in the previous section, we will work with a charged scalar field, obviously all results 
being immediately adaptable also to the case of neutral scalar field, in a straightforward way. 
In the charged field two distinct annihilation operator types appear: the $a$-type for particles and the 
$b$-type for antiparticles in the initial state field modes (and $c$-type and $d$-type for the outgoing field modes).
In line of principles, we could consider a initial coherent state constructed as the tensor product of 
$|\eta^{IN}>$ and $|\bar{\eta}^{IN}>$, 
where 
\beqnl
a_i^{IN} |\eta^{IN}> &=& \eta_i |\eta^{IN}>,\\
b_i^{IN} |\bar{\eta}^{IN}> &=& \bar{\eta}_i |\bar{\eta}^{IN}>,
\eeqnl
(we have omitted the suffix $IN$ where no confusion is possible), 
so that the initial state is simply 
\beq
|\zeta^{IN}>=|\eta^{IN}>\otimes |\bar{\eta}^{IN}>.
\eeq
We get
\beqnl
<\zeta^{IN}| N_k^{OUT} |\zeta^{IN}>&=& \sum_j |\beta_{kj}|^2+
\sum_{jl}  (\alpha_{kj} \alpha_{kl}^\ast  \eta_j^\ast \eta_l   
-\alpha_{kj}  \beta_{kl}^\ast   \eta_j^\ast \bar{\eta}_l^\ast \cr    
&&-\beta_{kj} \alpha_{kl}^\ast   \bar{\eta}_j \eta_l                  
+\beta_{kj}  \beta_{kl}^\ast   \bar{\eta}_j \bar{\eta}_l^\ast).   
\eeqnl
Spontaneous pair creation contribution is the first term on the right side, and is 
the usual one occurring in absence of incoming particles/antiparticles. The remaining terms 
are stimulated contributions. Analogously, 
\beqnl
<\zeta^{IN}| \bar{N}_k^{OUT} |\zeta^{IN}> 
&=& \sum_j |\beta_{kj}|^2 
+\sum_{jl}  (\alpha_{kj} \alpha_{kl}^\ast  \bar{\eta}_j^\ast  \bar{\eta}_l 
-\alpha_{kj}  \beta_{kl}^\ast  \bar{\eta}_j^\ast \eta_l^\ast  \cr                  
&& -\beta_{kj} \alpha_{kl}^\ast \eta_j \bar{\eta}_l                                      
+\beta_{kj}  \beta_{kl}^\ast \eta_j \eta_l^\ast).                                     
\eeqnl
We note that, in absence of initial antiparticle states, we obtain 
$|\zeta^{IN}>=|\eta^{IN}>\otimes |0>$, and then
\beq
<\zeta^{IN}| N_k^{OUT} |\zeta^{IN}>=\sum_j |\beta_{kj}|^2 +
\sum_{jl}  \alpha_{kj} \alpha_{kl}^\ast  \eta_j^\ast \eta_l   
\eeq
and
\beq
<\zeta^{IN}| \bar{N}_k^{OUT} |\zeta^{IN}> 
= \sum_j |\beta_{kj}|^2 
+\sum_{jl}  \beta_{kj}  \beta_{kl}^\ast \eta_j \eta_l^\ast.                                     
\eeq
With respect to the case where a Fock initial state is considered, we face with a situation 
where nondiagonal contributions appear.\\ 
In the diagonal case, we obtain 
\beqnl
<\zeta^{IN}| N_k^{OUT} |\zeta^{IN}>&=&|\beta_{k}|^2 +
|\alpha_{k}|^2  |\eta_k|^2 \\   
<\zeta^{IN}| \bar{N}_k^{OUT} |\zeta^{IN}> 
&=& |\beta_{k}|^2 
+|\beta_{k}|^2 |\eta_k|^2.                                     
\label{diag-coherent}
\eeqnl
By choosing the diagonal Bogoliubov transformation corresponding to the passage from Hartle-Hawking 
modes to Boulware ones thermality emerges again. It is also interesting to point out that (\ref{diag-coherent}) coincides 
with (\ref{diag-number}) under the same hypothesis of zero initial antiparticles, due to the fact that $|\eta_k|^2=<N_k^{IN}>$. 
In order to distinguish between the two different situations it is useful to calculate the variance  
\beq
(\delta^2 N_k^{OUT})_\varphi = <\varphi^{IN}| (N_k^{OUT})^2 |\varphi^{IN}>-<\varphi^{IN}| N_k^{OUT} |\varphi^{IN}>^2,
\eeq	
where $\varphi^{IN}$ is the initial state of interest.\\
A bit long calculation shows that 
variance is different in the two cases, and provides us a simple tool for discriminating between the aforementioned 
physical situations.\\
We first take into account the Fock space case. We note that the presence of a Bogoliubov transformation 
makes nontrivial the result, in the sense that, even if trivially we have 
$<\psi_{IN}|(N_k^{IN})^2 |\psi_{IN}>-<\psi_{IN}|N_k^{IN} |\psi_{IN}>^2=0$,  
 one obtains  
\beqnl
(\delta^2 N_k^{OUT})_\psi&=&<\psi^{IN}| (N_k^{OUT})^2 |\psi^{IN}>-<\psi^{IN}| N_k^{OUT} |\psi^{IN}>^2\cr
&=&\sum_{ij} \left[|\alpha_{ki}|^2 |\alpha_{kj}|^2 (<n_i^{IN}>+1)<n_j^{IN}>\right.\cr
&&\left. + |\beta_{ki}|^2 |\beta_{kj}|^2 <\bar{n}_i^{IN}> (<\bar{n}_j^{IN}>+1) \right.\cr
&& \left. +     |\beta_{ki}|^2    |\alpha_{kj}|^2    <n_i^{IN}>   <\bar{n}_j^{IN}> \right. \cr
&&\left. +  |\alpha_{ki}|^2    |\beta_{kj}|^2  (<n_i^{IN}>+1)  (<\bar{n}_j^{IN}>+1)\right].
\eeqnl
In absence of initial antiparticle states we find 
\beqnl
(\delta^2 N_k^{OUT})_\psi&=&
\sum_{ij} \left[|\alpha_{ki}|^2 |\alpha_{kj}|^2 (<n_i^{IN}>+1)<n_j^{IN}>\right. \cr
&&\left. +
                            |\alpha_{ki}|^2    |\beta_{kj}|^2  (<n_i^{IN}>+1)\right].
\eeqnl
In the diagonal case one obtains 
\beqnl
(\delta^2 N_k^{OUT})_\psi&=
|\alpha_{k}|^4 (<n_k^{IN}>+1)<n_k^{IN}>+
                            |\alpha_{k}|^2    |\beta_{k}|^2  (<n_k^{IN}>+1).
\eeqnl
In the coherent state representation we have a non vanishing variance on the number of outgoing particles too. A direct computation gives 
\beqnl
(\delta^2 N_k^{OUT})_\zeta&=&<\zeta^{IN}| (N_k^{OUT})^2 |\zeta^{IN}>-<\zeta^{IN}| N_k^{OUT} |\zeta^{IN}>^2\cr
&=&\sum_j |\alpha_{kj}|^2 \sum_i |\beta_{ki}|^2\cr
&&+[ \sum_j (|\alpha_{kj}|^2+|\beta_{kj}|^2) ]\ |\sum_i (\alpha_{ki}\eta^*_i-\beta_{ki}\bar \eta_i)|^2.
\eeqnl
In absence of initial antiparticle states it becomes
\beqnl
(\delta^2 N_k^{OUT})_\zeta&=\sum_j |\alpha_{kj}|^2 \sum_i |\beta_{ki}|^2+[ \sum_j (|\alpha_{kj}|^2+|\beta_{kj}|^2) ]\ |\sum_i \alpha_{ki}\eta^*_i|^2.
\eeqnl
In the diagonal case, we get
\begin{eqnarray}
(\delta^2 N_k^{OUT})_\zeta&=|\alpha_{k}|^2 |\beta_{k}|^2+ (|\alpha_{k}|^2+|\beta_{k}|^2) \ |\alpha_{k}|^2 |\eta_k|^2.
\end{eqnarray}
Summarizing, in absence of initial antiparticles and in the diagonal case, we find in the Fock state case
\beqnl
(\delta^2 N_k^{OUT})_\psi&=&
|\alpha_{k}|^4 <n_k^{IN}>^2 +|\alpha_{k}|^4 <n_k^{IN}>\cr
&&+
                            |\alpha_{k}|^2    |\beta_{k}|^2  <n_k^{IN}>+|\alpha_{k}|^2    |\beta_{k}|^2,
\eeqnl
to be compared with the coherent state case
\begin{eqnarray}
(\delta^2 N_k^{OUT})_\zeta&=|\alpha_{k}|^4 <n_k^{IN}>+|\beta_{k}|^2\ |\alpha_{k}|^2 <n_k^{IN}>+
|\alpha_{k}|^2 |\beta_{k}|^2.
\end{eqnarray}
Then we find 
\beq
(\delta^2 N_k^{OUT})_\psi-(\delta^2 N_k^{OUT})_\zeta=|\alpha_{k}|^4 <n_k^{IN}>^2.
\eeq
As a consequence, variance is a good tool for discriminating between the Fock state case and the coherent state 
one, when measurements of the expectation value of $N_k^{OUT}$ is not decisive, as shown above.

\section{Stimulated emission and the electromagnetic field}

The presence of the electromagnetic field is made more involved because of the gauge invariance, 
requiring a suitable gauge fixing, and the appearance, in the case e.g. of a covariant gauge, of 
spurious degrees of freedom (scalar component and longitudinal component of the field) to be 
suitably taken into account. We do not delve into a detailed discussion (for the quantization of the 
electromagnetic field on a manifold and in black hole backgrounds see e.g. \cite{dimock,crispino,pfenning}), we limit ourselves to 
some basic considerations which still shed light on stimulated emission in the case of 
the electromagnetic field.\\
Our key-ansatz consists in assuming that, e.g. in a generalized Feynman gauge \cite{crispino} 
it is possible to separate two mutually orthogonal physical polarizations $\lambda=1,2$ in such a way that they remain 
orthogonal to scalar and longitudinal polarizations. This is ensured in the case of dielectric black holes, 
because asymptotic states are well-known polariton states of standard flat spacetime \cite{PRD2015,physicascripta,exact}. 
The polarization index appears as a further label to be 
considered together with quantum number specifying solutions of the homogeneous Maxwell equations in presence of a black hole. 
In this sense, there are no substantial changes with respect to the (neutral) scalar field case, for what concerns 
the calculation of the Bogoliubov coefficients. 
Indeed, one has e.g. 
\beq
\beta_{\lambda k,\lambda' k'} = - (p_\mu^{\lambda' k'},f_\mu^{\lambda k \ast}),
\eeq
in a straightforward generalization of the scalar field case. $k$ stays for the set of labels specifying 
solutions of the Maxwell equations, and it is interesting to stress that we are only interested to the cases where 
$\lambda,\lambda'=1,2$, i.e. only physical polarizations have to be taken into account in order to 
calculate physical observables. Indeed, the remaining polarizations appear only in the set of unphysical variables, 
and do not participate any physically relevant number operator.  Then we obtain 
\beqnl
< \psi_{IN} | N_{\lambda k}^{OUT} |\psi_{IN} > &=& \sum_{\lambda'j} |\beta_{\lambda k,\lambda' j}|^2 + \sum_{\lambda'j} 
|\alpha_{\lambda k,\lambda' j}|^2 <n_{\lambda' j}^{IN}>\cr 
&&+ 
 \sum_{\lambda' j} |\beta_{\lambda k,\lambda' j}|^2 <\bar{n}_{\lambda' j}^{IN}>, 
\eeqnl
where $\lambda,\lambda'=1,2$. One can obtain straightforwardly an analogous equation for $\bar{N}_{\lambda k}^{OUT}$. 
In particular, in the thermal particle creation case one finds for the physical polarizations
\beqnl
< \psi_{IN} | N_{\lambda \omega}^{OUT} |\psi_{IN} > &=& \frac{1}{e^{\beta \omega}-1} +   
\frac{e^{\beta \omega}}{e^{\beta \omega}-1}<n_{\lambda \omega}^{IN}> \cr
&&+ 
\frac{1}{e^{\beta \omega}-1}  <\bar{n}_{\lambda \omega}^{IN}>,\\ 
< \psi_{IN} | \bar{N}_{\lambda \omega}^{OUT} |\psi_{IN} > &=& \frac{1}{e^{\beta \omega}-1} +   \frac{e^{\beta \omega}}{e^{\beta \omega}-1}
<\bar{n}_{\lambda \omega}^{IN}>\cr 
&&+ 
\frac{1}{e^{\beta \omega}-1}  <n_{\lambda \omega}^{IN}>. 
\eeqnl
As expected, the polarization does not affect spontaneous emission, which is clearly a unpolarized contribution. 
The two further terms, instead, depend on $\lambda$ as far as the initial particle and/or antiparticle states are 
polarized. If we assume that zero antiparticles are present in the initial state, then, 
\beqnl
< \psi_{IN} | N_{\lambda \omega}^{OUT} |\psi_{IN} > &=& \frac{1}{e^{\beta \omega}-1} +   
<n_{\lambda \omega}^{IN}>+\frac{1}{e^{\beta \omega}-1}<n_{\lambda \omega}^{IN}> , \label{enne-p}\\ 
< \psi_{IN} | \bar{N}_{\lambda \omega}^{OUT} |\psi_{IN} > &=& \frac{1}{e^{\beta \omega}-1} +
\frac{1}{e^{\beta \omega}-1}  <n_{\lambda \omega}^{IN}> \label{enne-n}. 
\eeqnl
The first contribution on the right side of both the above equations is the spontaneous (unpolarized) contribution. The second 
one in (\ref{enne-p}) is the number of photons in the IN state, and the third one in the same equation 
is the stimulated (polarized) contribution. In (\ref{enne-p}) we find, beyond the spontaneous contribution, a further 
polarized stimulated term.  In the limit for 
$n_{\lambda \omega}^{IN}\gg 1$, one finds 
\beqnl
< \psi_{IN} | N_{\lambda \omega}^{OUT} |\psi_{IN} > &\sim&  <n_{\lambda \omega}^{IN}>+
\frac{1}{e^{\beta \omega}-1}<n_{\lambda \omega}^{IN}>,\\ 
< \psi_{IN} | \bar{N}_{\lambda \omega}^{OUT} |\psi_{IN} > &\sim& \frac{1}{e^{\beta \omega}-1}  <n_{\lambda \omega}^{IN}>. 
\eeqnl
and 
\beq
< \psi_{IN} | \bar{N}_{\lambda \omega}^{OUT} |\psi_{IN} >\sim < \psi_{IN} | N_{\lambda \omega}^{OUT} |\psi_{IN} > 
\eeq
only if $e^{\beta \omega}-1\ll 1$.\\

It is straightforward to generalize to the electromagnetic case our analysis involving initial coherent states. 
We limit ourselves to the general formula for physical polarizations:
\beqnl
<\zeta^{IN}| N_{\lambda k}^{OUT} |\zeta^{IN}>&=& \sum_{\lambda' j} |\beta_{\lambda k,\lambda' j}|^2+
\sum_{\lambda' j,\lambda'' l}  (\alpha_{\lambda k, \lambda' j} \alpha_{\lambda k, \lambda'' l}^\ast  \eta_{\lambda' j}^\ast \eta_{\lambda'' l} \cr  
&&-\alpha_{\lambda k, \lambda' j}  \beta_{\lambda k, \lambda'' l}^\ast   \eta_{\lambda' j}^\ast \bar{\eta}_{\lambda'' l}^\ast     
-\beta_{\lambda k, \lambda' j} \alpha_{\lambda k, \lambda'' l}^\ast   \bar{\eta}_{\lambda' j} \eta_{\lambda'' l}  \cr                
&&+\beta_{\lambda k, \lambda' j}  \beta_{\lambda k, \lambda'' l}^\ast   \bar{\eta}_{\lambda' j} \bar{\eta}_{\lambda'' l}^\ast).   
\eeqnl
Analogously, one can obtain  $<\zeta^{IN}| \bar{N}_{\lambda k}^{OUT} |\zeta^{IN}> $. 
In absence of initial antiparticle states one gets
\beq
<\zeta^{IN}| N_{\lambda k}^{OUT} |\zeta^{IN}>=\sum_{\lambda' j} |\beta_{\lambda k,\lambda' j}|^2 +
\sum_{\lambda' j,\lambda'' l}  \alpha_{\lambda k, \lambda' j} \alpha_{\lambda k, \lambda'' l}^\ast  \eta_{\lambda' j}^\ast \eta_{\lambda'' l},     
\eeq
which 
in the diagonal case becomes 
\beq
<\zeta^{IN}| N_{\lambda k}^{OUT} |\zeta^{IN}>=|\beta_{\lambda k}|^2 +
|\alpha_{\lambda k}|^2  |\eta_{\lambda k}|^2.          
\eeq
It is also straightforward to adapt our previous analysis concerning the variance to the electromagnetic field case.

\begin{figure}[htbp] 
\centering
\includegraphics[angle=0,width=8cm]{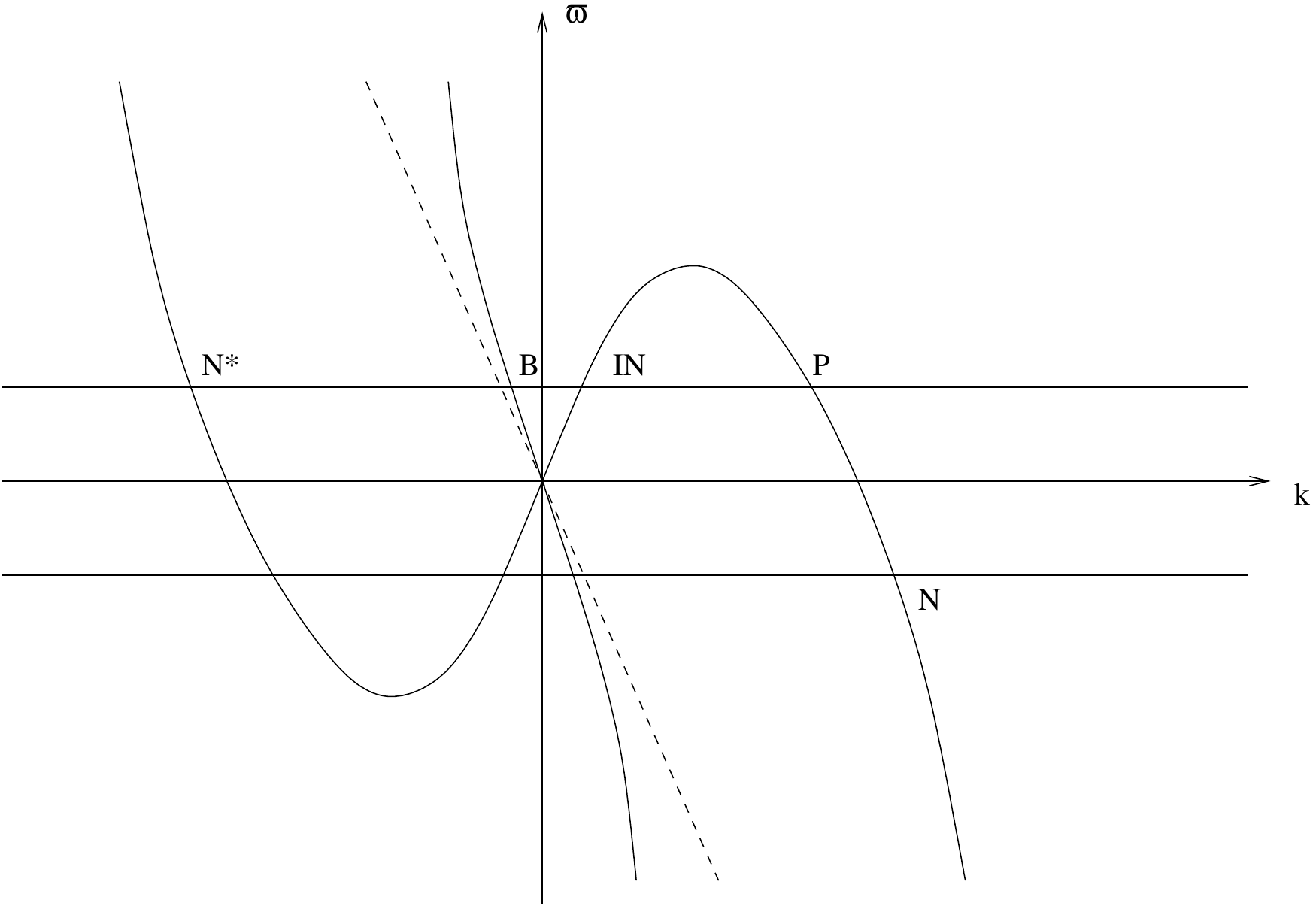}
	\caption{ \label{fig:fig1} Asymptotic dispersion relation for the Cauchy approximation in the comoving frame. The dashed line divides antiparticle states (below it) and particle ones (above it). $\omega$ and $k$ are the frequency and the wave number in the comoving frame respectively. Two lines at $\omega=$const and at $-\omega=$const are
	also drawn, and relevant states are explicitly  indicated.}
	\label{fig:asymptDR}
\end{figure}

In the case of stimulated 
Hawking effect in dielectric media the ternary process $IN\to P+N$ takes place, 
where $IN$ stays for the initial state mode (incoming particle), $P$ labels the outgoing particle mode, and 
$N$ the outgoing antiparticle mode  \cite{rubino-njp,petev-prl,PRD2015}. In figure \ref{fig:asymptDR} we display the typical 
situation in the case of the so-called Cauchy approximation for the Sellmeier equation for a single resonance. We also point out that we work in the comoving frame, i.e. in the frame of the laser pulse. 
There is a monotone branch, which provides a state $B$ which decouples from the spectrum in the case of not too stiff 
refractive index variation (see e.g. \cite{PRD2015}), and a non-monotone branch which is instead deeply involved in the analogue 
Hawking effect and provides three states $IN,P,N$. 

We stress that, in the comoving frame of 
the dielectric perturbation, we can adopt as asymptotic states the unperturbed states of the covariant Hopfield model 
we set up in \cite{physicascripta,exact}. In this case, physical polarizations are found in the framework we assumed above 
(i.e. they are actually mutually orthogonal, and, in particular, orthogonal to the longitudinal and scalar polarizations, which 
do not participate to the physical states).\\
Let us interpret the above results in terms of the observed $P$-peak (particle peak) and $N$-peak (antiparticle peak).  Our considerations are limited to the case where 
thermality occurs and no relevant contribution to the scattering process of the fourth mode involved in the process, called  $B$-mode,  appears. (\ref{enne-p}) represents the 
mean number of photons in the $P$-peak, whereas (\ref{enne-n}) stays for the number of antiparticles in the $N$-peak. 
As it is evident, in the $P$-peak we find the initial state photons (i.e. photons in the state $IN$), and, as an effect of the thermal 
particle creation, of the stimulated and spontaneous created particles. In the $N$-peak, we find 
stimulated and spontaneous created antiparticles. We have that the pair-created photons participate both the peaks, 
the particle modes joining the $P$-peak together with the $IN$-photons, and the antiparticle modes forming the $N$-peak, 
which represents a clear signal of amplification (pair creation). As to the polarization, we note that both the peaks, as far as 
the stimulated contribution becomes the leading one, are polarized. In particular, polarized $IN$-photons generate particles and antiparticles with the same polarization as for the $IN$-photons.\footnote{Obviously particles and
antiparticles initially have opposite phases, but in terms of experimental observation this fact is irrelevant since in general the phases will change independently along the story of the single particle.}

\section{Conclusions}

We have reconsidered stimulated emission in a black hole context, in light of the fact that analogue gravity 
allows a relevant role for stimulated emission, in spite of its substantial practical irrelevance for astrophysical black holes. 
We have recalled existing results for black holes, and we have pointed out as, in the case of Hartle-Hawking state, 
stimulated emission can be calculated in a thermofield dynamics framework. Then, we have considered coherent states 
as possible initial states, in view of the possibility in a dielectric black hole case to stimulate Hawking-like emission 
by means of a laser probe. Then we have also extended the results to the case of electromagnetic field, which, under 
reasonable hypotheses, are (non-trivial) extensions of the standard scalar field results.\\
Finally, we have applied our analysis to the dielectric black hole case. The created pairs are such that one emitted photon  is found in the 
P mode peak and the companion photon (antiparticle) is found in the N mode peak. The spontaneous contribution 
is unpolarized, as it should be due to thermality of the spontaneous radiation, whereas the stimulated one is 
suitably polarized, in such a way that a created photon is polarized in the same way as the stimulating particle state, 
and is polarized in the opposite way as the stimulating antiparticle state. As a consequence, given an IN state 
populated by particle states, one obtains a P mode peak and an N mode peak, all having the same polarization.

\newpage

\end{document}